\documentclass[preprints,article,accept,moreauthors,pdftex]{Definitions/mdpi} 
\renewcommand{\linenumbers}{}
\firstpage{1} 
\pubvolume{1}
\issuenum{1}
\articlenumber{0}
\pubyear{2021}
\copyrightyear{2020}
\datereceived{} 
\dateaccepted{} 
\datepublished{} 
\hreflink{https://doi.org/} 
\pdfoutput=1

\Title{Multi-modal Residual Perceptron Network for Audio-Video Emotion Recognition}

\TitleCitation{Multi-modal Residual Perceptron Network for Audio-Video Emotion classification}

\Author{Xin Chang, W\l adys\l aw Skarbek}

\AuthorNames{Xin Chang and W\l adys\l aw Skarbek}

\AuthorCitation{Chang, X.; Skarbek, W.}

\address{%
Warsaw University of Technology, Institute of Radioelectronics and Multimedia Technology}

\corres{xin.chang.dokt@pw.edu.pl}

\abstract{Emotion recognition is an important research field for Human-Computer Interaction.
Audio-Video Emotion Recognition is now attacked with Deep Neural Network modeling tools. In published papers, as a rule, the authors show only cases of the superiority in multi-modality over audio-only or video-only modality. However, there are cases superiority in uni-modality can be found. In our research, we hypothesize that for fuzzy categories of emotional events, the within-modal and inter-modal noisy information represented indirectly in the parameters of the modeling neural network impedes better performance in the existing late fusion and end-to-end multi-modal network training strategies. To take advantage and overcome the deficiencies in both solutions, we define a Multi-modal Residual Perceptron Network which performs end-to-end learning from multi-modal network branches, generalizing better multi-modal feature representation. For the proposed Multi-modal Residual Perceptron Network and the novel time augmentation for streaming digital movies, the state-of-art average recognition rate was improved to 91.4\% for The Ryerson Audio-Visual Database of Emotional Speech and Song dataset and to 83.15\% for Crowd-sourced Emotional multi-modal Actors dataset. Moreover, the Multi-modal Residual Perceptron Network concept shows its potential for multi-modal applications dealing with signal sources not only of optical and acoustical types.
} 


\begin{document}

\begin{figure}[ht]
\begin{center}
\includegraphics[width=0.70\columnwidth]{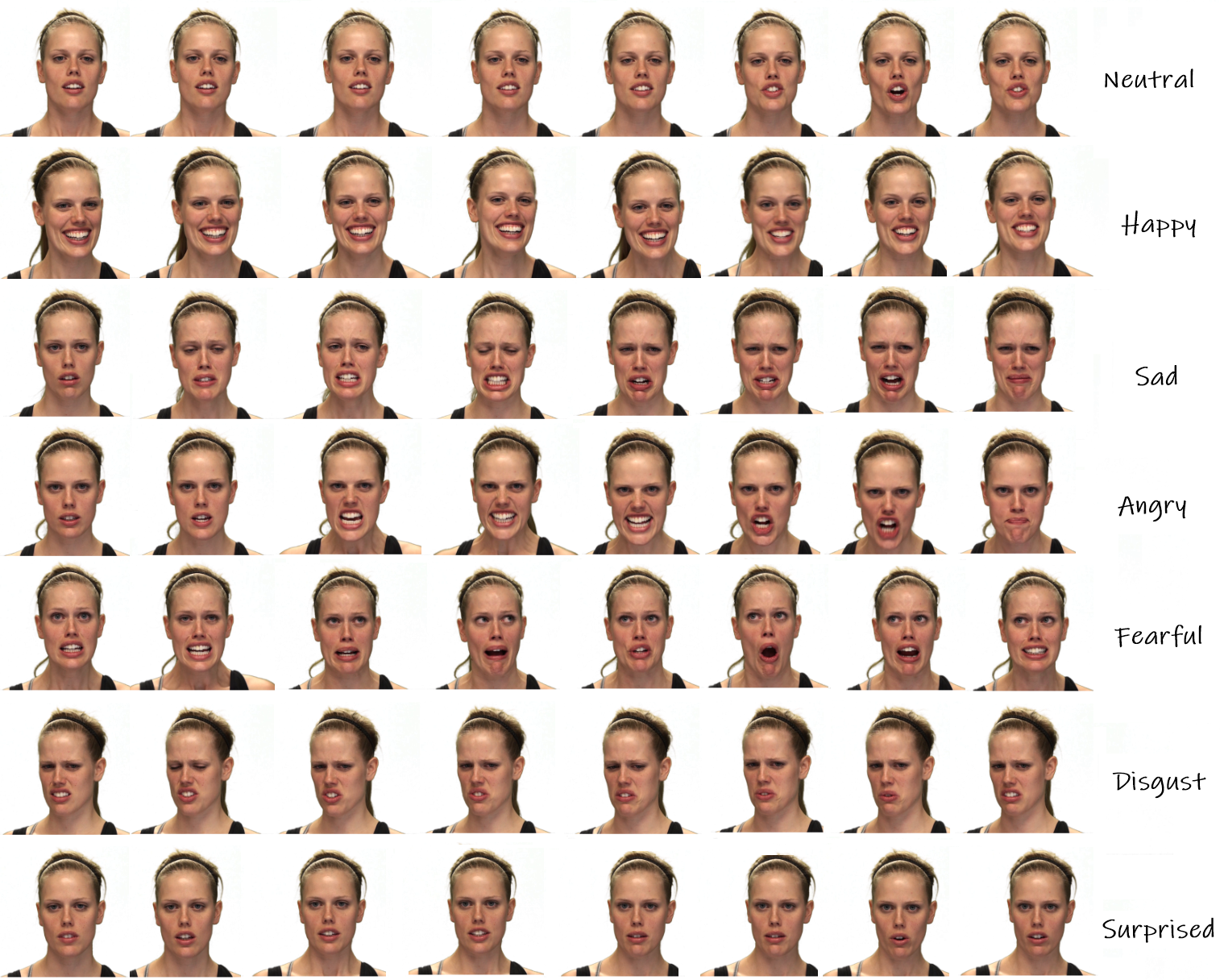}
\caption{Video frames of visual facial expressions selected from RAVDESS (Ryerson Audio-Visual Database of Emotional Speech and Song) dataset.}
\label{visual_emo_frames}
\end{center}
\end{figure}

\begin{figure}[ht]
\begin{center}
\includegraphics[width=1.0\columnwidth]{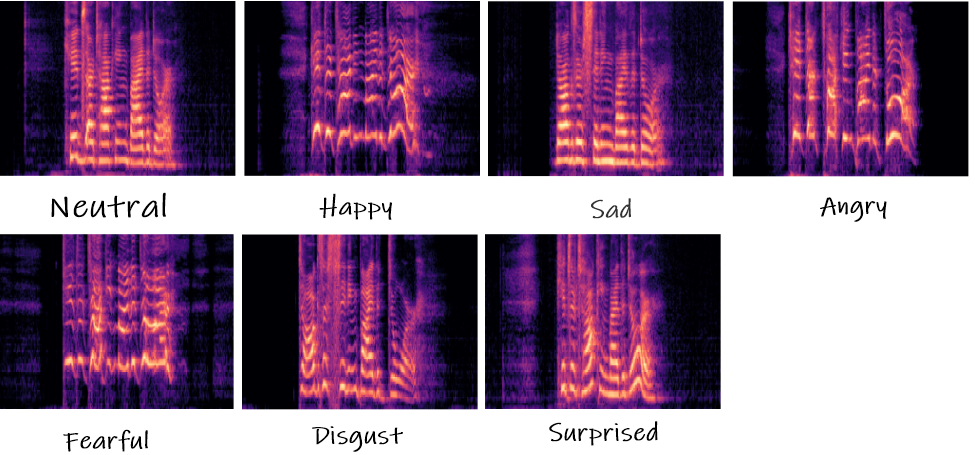}
\caption{Mel Spectrograms of vocal timbres selected from RAVDESS (Ryerson Audio-Visual Database of Emotional Speech and Song) dataset.}
\label{vocal_emo_frames}
\end{center}
\end{figure}

\end{paracol}
\begin{figure}[H]
\begin{center}
\includegraphics[width=.7\columnwidth]{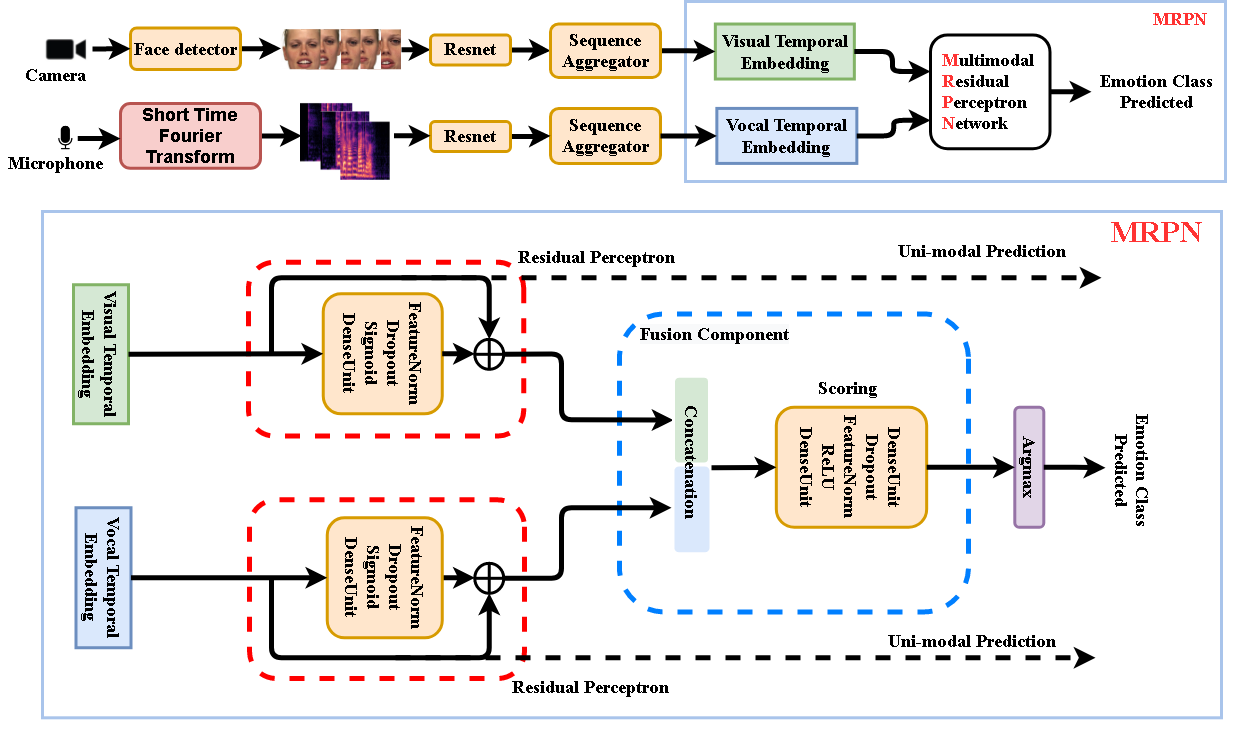}
\caption{The proposed multi-modal emotion recognition system using Deep Neural Network (DNN) approach. \textbf{ Upper part:} Video frames and audio spectral segments get independent temporal embeddings to be fused by our multi-modal Residual Perceptron Network (MRPN). \textbf{ Lower part:} MRPN performs in each modality normalizations via the proposed Residual Perceptrons and then scores their concatenated outputs in the Fusion Component. The uni-modal prediction branches are only active in training mode.}
\label{briefdesign}
\end{center}
\end{figure}
\begin{paracol}{2}
\linenumbers
\switchcolumn

\section{Introduction}\label{intro}
This paper presents a novel end-to-end 
Deep Neural Network (DNN) based framework, as Figure~\ref{briefdesign} illustrates, addressing the Audio-Video Emotion Recognition (AVER) problem. Just like human beings understand emotional expressions in our daily social activates through multi-senses (e.g., visual, vocal, textually meaningful), 
neural computing units as part of intelligent artificial sensors are now playing important roles in emotion recognition tasks. Specialized sensors in Human-Computer Interaction (HCI) capture information responsible for understanding visual and vocal information just like we -- the human beings -- understand emotional expressions through our multi-senses.

\subsection{Emotion recognition from face expression and voice timbre}
Intuitively, functionalities of intelligent artificial neurons are assigned with similar concepts as our brain cells are, processing information from the very raw senses independently and appropriately to their types. 

Visually, information captured by the camera is distributed to several frames as Figure~\ref{visual_emo_frames} shows. Discrete information in a single frame is firstly delivered to pattern extracting intelligent sensors for features such as Fisherfaces and Eigenfaces \cite{10.1109/34.598228} or the deep features from Convolution Neural Network (CNN) \cite{Lecun98gradient-basedlearning}. To fully preserve the information from the discrete signals, some Sequence Aggregation Component (SAC), e.g., Long Short Term Memory (LSTM) \cite{HochSchm97} or Transformer \cite{vaswani2017attention}, is then needed to further process the extracted features. Finally, a classifier such as Support Vector Machine (SVM) or some neural dense layer takes the integrated features for the classification. 

\end{paracol}
\begin{paracol}{2}
\linenumbers
\switchcolumn

Raw vocal inputs are usually with 10,000 to 44,100 samples per second, while the visual frame rate is about 25-30 image frames per second. Though raw digital signals in the time domain approximate the original signal precisely, their spectral representation, e.g., Spectrogram frame, Mel-spectrogram coefficients, or Log Mel-spectrogram frame, appeared more effectively for sound recognition. The spectral converted vocal signals have shown significant improvements in many classification problems, in spite of some limitations.
Expression events do not last at the same time, thus the width of the Spectrogram frames changes, which is not desirable for CNN pattern extractors. 
Therefore, the extracted features also need further processing from SAC which outputs integrated features.
Figure~\ref{vocal_emo_frames} shows the expression events from different categories and with different time duration.

\end{paracol}
\begin{paracol}{2}
\linenumbers
\switchcolumn
\subsection{Multi-modal emotion recognition}
AVER solution also follows the sensation of human beings, people claim they hear the sound when looking at the sheet music, smell the odor when recalling the memory from a photo or see the sea sight from the smell of the air. The multi-sensation information is processed by different areas of our cerebral cortex, movement, hearing, seeing, etc., then highly correlated by some other brain areas. 
Thus, the decision made depends not just on the recognition of uni-modal sensations independently but also jointly.

The learning process of neural sensors should also mimic our learning process. The neurons shape their weights just like our cerebral cortex changes from the stimulation of the environments and look for any correlation of them during the learning process in the supervised neural network sensors training. However, we claim the existing late fusion and end-to-end training strategies hold their own advantages but also deficiencies.

\subsection{Paper contribution and structure}
This paper shows by experiments the deficiencies of two training strategies: late fusion and end-to-end. Late fusion strategy takes trained static uni-modal networks and trains their fusing network components. End-to-end strategy trains all the multi-modal and uni-modal components together. A novel architecture is proposed to take advantage of both solutions and avoid their side effects, respectively. We demonstrate superiority in the novel end-to-end mechanism and architecture comparing with naive fusion mechanism in either late fusion or end-to-end training. The proposed DNN framework, data augmentation procedures, and network optimization strategy are discussed. The detailed analysis and discussion are presented by computing experiments on The Ryerson Audio-Visual Database of Emotional Speech and Song (RAVDESS) and Crowd-Sourced Emotional Multimodal
Actors Dataset (Crema-d) datasets. Our major contributions are concluded as follows:
\begin{enumerate}
\item \textit{Multi-modal framework:} We propose a novel within-modality Residual Perceptrons (RP) for efficient gradient blending for the neural network optimization using multi-term loss function in MRPN. The sub-networks and target loss functions produce superior parameterized multi-modal features, preserving the original knowledge of uni-modalities, which impedes inter-modal learning. The within-modality RP components reduce the side effects brought from such multi-term loss functions. As the result, we got significantly better performance over direct strategies including late fusion and end-to-end without MRPN.
\item \textit{Time Augmentation of input frames:} We demonstrate data augmentation in time involving randomly slicing over input frame sequences from both modalities improved the recognition performance to the state-of-art, even without MRPN. We show also the results that time augmentation doesn't solve the cases where uni-modal solutions are better than multi-modal solutions, yet solved by MRPN.
\end{enumerate}
\section{Related work}
\subsection{Superiority in multi-modal approach}
Many have shown significant improvement of multi-modal solutions. N. Neverova et al. \cite{neverova2015moddrop} suggest gradual fusion involving the random dropping of separate channels, and this method was adopted by V. Vielzeuf et al. \cite{vielzeuf2017temporal} in the AVER solution for their best result.

Fusion at the early or late stage is discussed by others. R. Beard et al. \cite{beard-etal-2018-multi} proposed multi-modal feature fusion at the late stage, while E. Ghaleb et al. \cite{8925444} try to project features to a shared space in the early stage and provided external loss functions to minimize the distance of features from different modalities. A. Zadeh et al. \cite{zadeh2018memory} proposed Multi-view Gated Memory to gate the multi-modal knowledge from LSTM in the time series. E. Mansouri-Benssassi and J. Ye \cite{8852473} archive early fusion by creating distinct multi-modal neuron groups.

S. Zhang et al. \cite{10.1109/TCSVT.2017.2719043} take features from CNN and 3D-CNN models for vocal and visual sources then make global averaging as video features. NC. Ristea et al. \cite{8906538} take the features extracted by CNNs from both modalities and exploit the fused features for classification purposes. E. Tzinis et al. \cite{tzinis2021improving} take cross-modal and self-attention modules. Y. Wu et al. \cite{Wu_2019_ICCV} localize events crossing modalities. E. Ghaleb et al. \cite{8935376} suggest multi-modal emotion recognition metric learning to create a robust representation for both modalities.

\subsection{Potential failures in the existing solutions}\label{cross-in}
Let's consider the human brain learning process again. Say the information is wrong in some of the sensory stimulation. A child learned an animal looks just like a dog but having the sound of the cat from the manipulated movies and this kid has never learned the dog and cat in a real-life environment. He will either see a dog and tell it's a cat or hear the cat sound and tell it's a dog. The situation can go even worse if the stimulation he learned from is also fuzzy within their own sensation.

His recognition is still intact to some extent that he can sometimes correctly recognize the visual or acoustic information pattern. But the recognized information is distorted, along with the correlation of the inter-modal information. This made the distorted uni-modal knowledge he learned having also a negative impact on the other. The same concept we address to the current multi-modal neural network solutions. The within-modal and inter-modal noisiness of the learned pattern both contribute to the wrong recognition.
Despite many advantages from the multi-modal solutions which boost the recognition performance of emotion recognition tasks, we hypothesize the uncontrolled fusion strategy, adopted by
\cite{vielzeuf2017temporal, Wu_2019_ICCV, 7945502, HOSSAIN201969, article, 10.1109/TCSVT.2017.2719043, beard-etal-2018-multi, zadeh2018memory} could lead to potential deficiencies in either late fusion or end-to-end training strategy. 

Though many have shown superior performance of late fusion strategy \cite{wang2016temporal, simonyan2014twostream, carreira2018quo}, for instance, for audio events detection in video material,
W. Wang et al. \cite{9156420} illustrate the results of naive fusion from multi-modal features can be worse than the best uni-modal approach. They propose blending the gradient flow by multi-task loss functions, which is referred to as multi-term loss function by us, from uni-modalities and multi-modality, which help better parameterization of the whole system in many other research areas. Though they suggest benefits from blending the gradient flows, multi-tasking could make the features hard to be optimized serving both uni-modal and multi-modal purposes suggested by many researchers \cite{standley2020tasks, 10.1023/A:1007379606734, Yu2020GradientSF}. We demonstrate how this proposal can still fail in some inferior cases but is solved by the within-modal RP component in MRPN.

\section{Hypothesis}
In this section, we discuss our hypothesis where fuzzy information from the uni-modalities can cause chaos in not just the uni-modal neurons but also the correlation neurons, namely the fusion component.
\subsection{Within-modal information can be missing or fuzzy}
The missing or fuzzy information can be noticed in either visual or vocal modality emotion recognition solution, and then the success rate of recognition can not be increased noticeably. Missing information refers to feature data where emotion categories are confused with neutral categories in the uni-modality. Fuzzy information stands for feature data where one emotion category cannot be distinguished from another one in the uni-modality.

Namely, the visual modality results from the challenge FER-2013 \cite{GOODFELLOW201559} for single image facial recognition have only improved about 4\% to 76.8\% over the past eight years by W. Wang et al. \cite{wang2020learning}.

Moreover, for video frames, HW. Ng et al. \cite{10.1145/2818346.2830593} got 47.3\% validation accuracy and 53.8\% testing accuracy on EmotiW dataset \cite{dhall2018emotiw} using transfer learning and averaged temporal deep features.
Similarly, for vocal solutions, the results for Interactive Emotional Dyadic Motion Capture
dataset (IEMOCAP) \cite{iemocap} with the raw inputs are reported around 76\% by S. Kwon \cite{s20010183} and 64.93\% by S. Latif et al. \cite{DBLP:journals/corr/abs-1712-08708}. The recognition rate for these cases is far from optimal.

Apart from the design, functionality, and training of the neural network, the human voting for those datasets draws our concerns. As the teacher in supervised learning, almost all datasets related to emotion recognition have unsatisfactory knowledge. The ones who understand human emotions the best, the human beings themselves, cannot make a majority of the agreement to the author's labeling. On average, the human rate of the emotional categories is 72\% from
 IEMOCAP, human accuracy on FER-2013 \cite{GOODFELLOW201559} is 65$\pm$5\%, Crema-d \cite{6849440} holds the accuracy of 63.6\% and RAVDESS \cite{10.1371/journal.pone.0196391} has the results of 72.3\%. 

All the reports point out that in every uni-modality, the information of data is never crystal, thus the learned knowledge of a uni-modality in emotion recognition, can be corrupted and uncontrolled by the network. We can't identify or agree on which samples are wrong because the boundaries of the clusters are quite subjective. 
\subsection{End-to-end modeling for multi-modal data can be distorted}\label{conflict}

Multi-modal solutions seem to find more generalized patterns via the extension of parameters.
However, the fused features have left a backdoor for distorted pattern learning, the side effects are concealed by its benefits.

\begin{figure}[ht]
\begin{center}
\includegraphics[width=0.9\columnwidth]{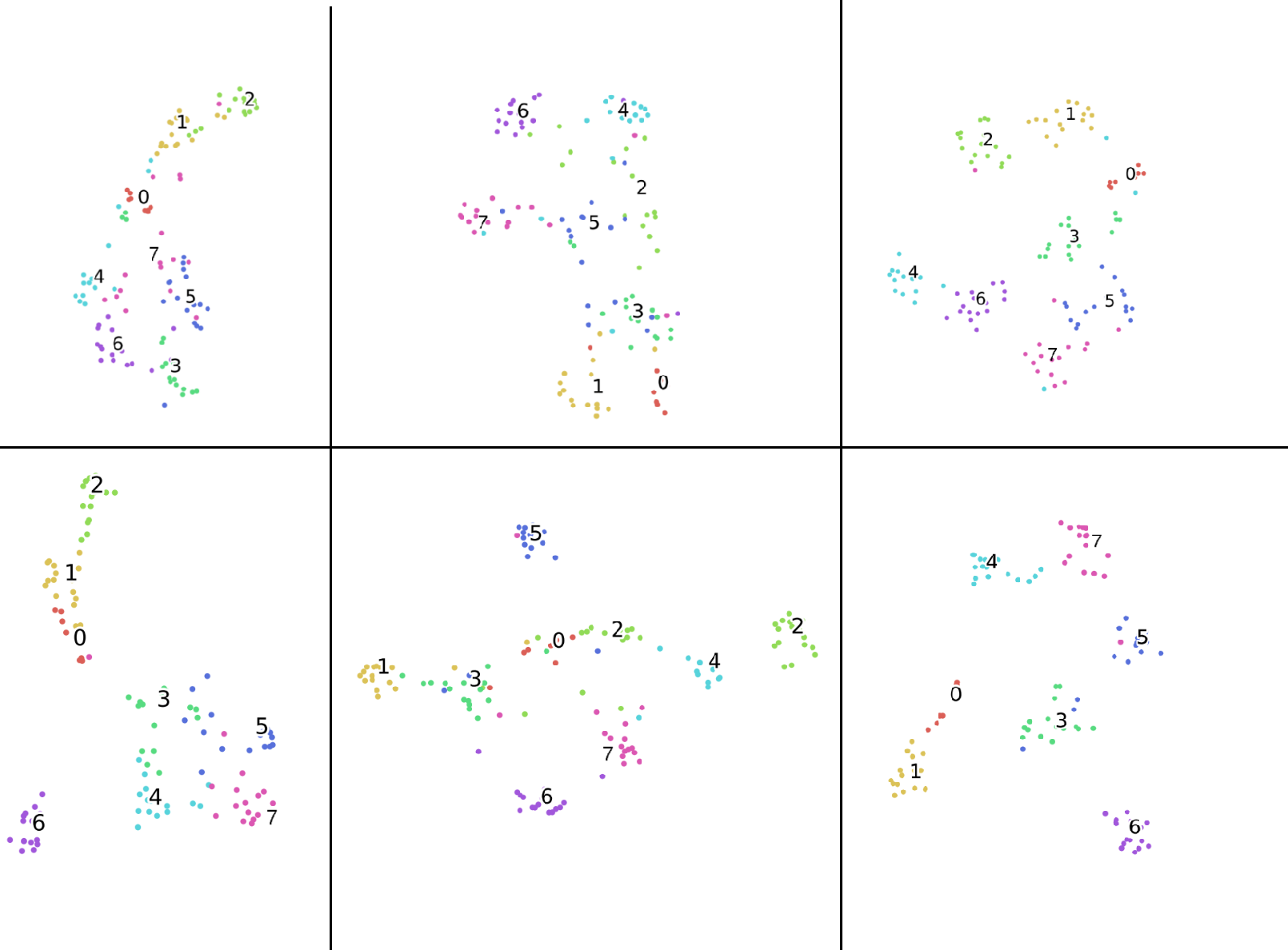}
\caption{Visualization (t-SNE algorithm) of deep features clustering from two different setups where the train/validation sets are shuffled. The clustering with respect to emotion classes are listed. 0: {\it neutral}, 1: {\it calm}, 2: {\it happy}, 3: {\it sad}, 4: {\it angry}, 5: {\it fearful}, 6: {\it disgust}, 7: {\it surprised}. \textbf{ Top part}: clustering results from one setup of uni-modalities and multi-modality. {\footnotesize \it Left part}: only image modality. {\footnotesize \it Middle part}: only audio modality. {\footnotesize \it  Right part}: multi-modality.\textbf{ Down part}: clustering results from the other setup where train/validation sets are shuffled.}
\label{cluster}
\end{center}
\end{figure}
Figure~\ref{cluster} shows us the different clustering from shuffled train/validation sub-datasets in each uni-modal solution and multi-modal solution, respectively. The actors in the validation set are different than they are in the training set.  Due to the missing and fuzzy information from the uni-modal data, the clustering of the same uni-modality shown in the top and bottom panels differs. It can be seen from the figure, for the same modality solutions, either from uni-modal or multi-modal, missing information is causing the overlapping clustering for the neutral category. The cases appear similarly for the overlapping of emotional categories caused by the fuzzy information. This suggests that patterns within uni-modality are hard to be generalized, aligned with human voting results mentioned.

Under such circumstances, we do not know which training sample is fuzzy in which modality, not just it causing the fuzzy direction of within-modal learning, but also inter-modal learning in end-to-end training. i.e. information in modality A is fuzzy while crystal in modality B, can results in correct learning for modality B yet fuzzy learning for modality A. In the end, the distribution of the wrong direction learned knowledge is unknown.

Figure~\ref{fusion_problem} illustrates the source of deficiency in the architecture during the gradient backpropagation, wherein the blue frame denotes the fusion component. Namely, the concatenation unit of the features from different modalities can backpropagate the gradients modifying jointly weights in each modality, potentially distort the knowledge in some modality. 
\end{paracol}
\begin{figure}[ht]
\begin{center}
\includegraphics[width=0.9\columnwidth]{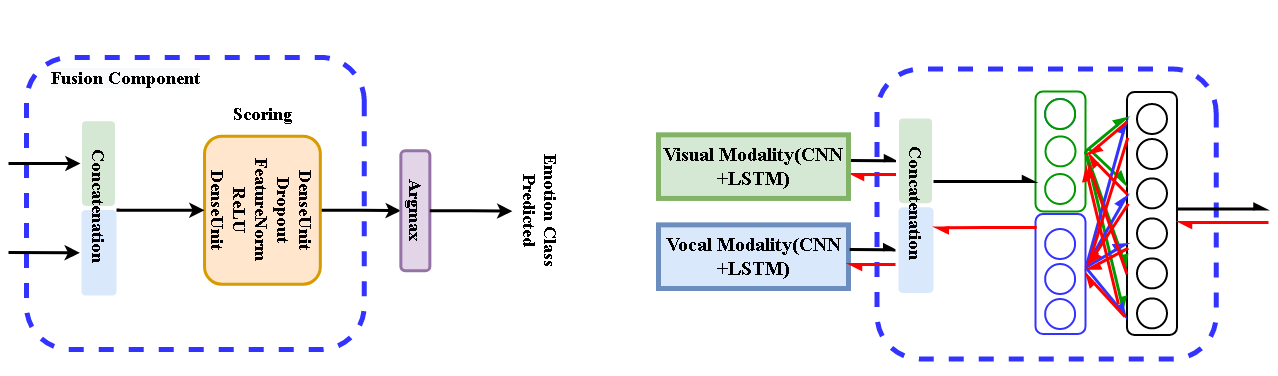}
\caption{Distorted gradients backpropagation in some modality since the gradients from fused layer makes impact on gradients flow into neural weights of both modalities.}
\label{fusion_problem}
\end{center}
\end{figure}
\begin{paracol}{2}
\linenumbers
\switchcolumn
\subsection{Late fusion modeling for multi-modal data can be insufficient}\label{conflict}
The late fusion seems to prevent the inter-modal learning of the system, however, not just the distribution of the fuzzy information to each modality is unknown, but also the clean data which holds the highly correlated information between modalities are. If the samples contain clean information in all modalities, then the frozen parameters of the shadow layers cannot make proper adjustments to learn inter-modal information from the joint gradient flow.

\section{Proposed Methods}
Addressing the mentioned issues, we proposed a novel MRPN along with a multi-term loss function for the better parameterization of the whole network taking advantage of both late fusion and end-to-end strategies while avoiding their deficiencies.  MRPN can eliminate the problems without assuming the data is noisy or clean.
\subsection{Functional description of analyzed networks}\label{functionality}

The functional descriptions of the analyzed deep networks are presented for their training mode
(see Figure~\ref{details}). They are based on the selected functionalities of neural units and components. We use index $m$ for inputs of any modality. In our experiments $m=v$ or $m=a$.
\begin{enumerate}
\item $F_m$: feature extractor for input temporal sequence $x_m$ of modality $m$, e.g. $F_v$ for video frames $x_v$, $F_a$ for audio segments $x_a$.

\item $A_m$: aggregation component SAC for temporal feature sequence leading to temporal feature vector $f_m$, eg. $A_v$, $A_a$ for video and audio features, respectively.

\begin{equation}
f_m \doteq A_m(F_m(x_m)) \longrightarrow
f_v \doteq A_v(F_v(x_v)),\  
f_a \doteq A_a(F_a(x_a))
\end{equation}
\item Standard computing units: {\tt DenseUnit} $-$ affine (a.k.a. dense, full connection), {\tt Dropout} $-$ random elements dropping for model regularizing, {\tt FeatureNorm} $-$ normalization for learned data regularizing (batch norm is adopted in the current implementation), and {\tt Concatenate} $-$ joining feature maps, {\tt ReLU, Sigmoid} $-$ activation units.

\item {\tt Scoring} $-$ component mapping feature vectors to class scores vector, usually composing the following operations:

\begin{equation}
\rightarrow DenseUnit 
\rightarrow ReLU
\rightarrow FeatureNorm
\rightarrow DenseUnit 
\end{equation}

\begin{equation}
\begin{array}{l}
m\in\{v,a\},\ \hat{f}_m\doteq FeatureNorm(f_m)
\longrightarrow s_m\doteq Scoring(\hat{f}_m)
\\[2pt]
g_{va} \doteq FeatureNorm(Concatenate(
g_{v},g_{a})) \longrightarrow
s_{va}\doteq Scoring(g_{va})\\
\end{array}
\end{equation}
\item {\tt FusionComponent} $-$ concatenates its inputs $g_v, g_a$, then makes the statistical normalization, and finally produces the vector of class scores:

\begin{equation}\label{MRPNfusion}
\begin{array}{l}
s_{va} \doteq FusionComponent(g_v,g_a)
\longrightarrow\\[2pt]
g_v, g_a\rightarrow Concatenate \rightarrow Scoring \rightarrow s_{va}
\end{array}
\end{equation}
In our networks $g_v, g_a$ are statistically normalized multi-modal features ($\hat{f}_v,\hat{f}_a$) or their residually updated form ($f'_v$, $f'_a$) -- cf. those symbols in Figure~\ref{details}. 

\item {\tt SoftMax} $-$ computing unit for normalization of class scores to class probabilities:

$$
\begin{array}{l}
m\in\{v,a\}\longrightarrow p_m \doteq Softmax(s_m)\\[2pt]
p_{va}\doteq Softmax(s_{va})
\end{array}
$$

\item {\tt CrossEntropy} $-$ a divergence of probability distributions used as loss function. Let $p$ is the target probability distribution. Then the following loss functions are defined:

\begin{equation}
\begin{array}{l}\label{MRPNloss}
m\in\{v,a\},\ p_m \doteq Softmax(s_m)
\longrightarrow \mathcal{L}_m \doteq CrossEntropy(p,p_m)\\[2pt]
p_{va}\doteq Softmax(s_{va}) \longrightarrow
\mathcal{L}_{va} = CrossEntropy(p,p_{va})\\[2pt]
\mathcal{L} \doteq \mathcal{L}_v + \mathcal{L}_a +
\mathcal{L}_{va}
\end{array}
\end{equation}
where $\mathcal{L}$ is {\it multi-term loss function} implying the gradient blending in the backpropagation stage. 
\item {\tt ResPerceptron} (Residual Perceptron) $-$ component performing statistical normalization for the dense unit (perceptron) computing residuals for normalized data. In our solution it transforms a modal feature vector $f_m$ into $f'_m$, as follows:

\begin{equation}\label{sigmoidact}
\begin{array}{l}
\hat{f}_m \doteq FeatureNorm(f_m) \longrightarrow
f'_m \doteq ResPerceptron(\hat{f}_m)
\longrightarrow\\[2pt]
f'_m \doteq\hat{f}_m + FeatureNorm(Sigmoid(DenseUnit(\hat{f}_m)))
\end{array}
\end{equation}

\end{enumerate}

Three networks $\mathcal{N}_0, \mathcal{N}_1, \mathcal{N}_2$ are defined for further analysis:
\begin{enumerate}
\item Network $\mathcal{N}_0(f_v,f_a;p)$ with fusion component and loss function $\mathcal{L}_{va}$:

\begin{equation}
\begin{array}{l}
\hat{f}_v\doteq FeatureNorm(f_v),\ 
\hat{f}_a\doteq FeatureNorm(f_a)\\[2pt]
s_{va} \doteq FusionComponent(\hat{f}_v,\hat{f}_a)
\\[2pt]
p_{va}\doteq SoftMax(s_{va}) \longrightarrow
\mathcal{L}_{va}\doteq CrossEntropy(p,p_{va})
\end{array}
\end{equation}

\item Network $\mathcal{N}_1(f_v,f_a;p)$ with fusion component and fused loss function 
$\mathcal{L} \doteq \mathcal{L}_v + \mathcal{L}_a +
\mathcal{L}_{va}$:

\begin{equation}
\begin{array}{l}\label{MRPNscores}
\hat{f}_v\doteq FeatureNorm(f_v),\ 
\hat{f}_a\doteq FeatureNorm(f_a)\\[2pt]
s_v\doteq DenseUnit(\hat{f}_v),\ 
s_a\doteq DenseUnit(\hat{f}_a),\ 
s_{va} \doteq FusionComponent(\hat{f}_v,\hat{f}_a)
\\[2pt]
p_{v}\doteq SoftMax(s_{v}) \longrightarrow
\mathcal{L}_{v}\doteq CrossEntropy(p,p_{v})
\\[2pt]
p_{a}\doteq SoftMax(s_{a}) \longrightarrow
\mathcal{L}_{a}\doteq CrossEntropy(p,p_{a})
\\[2pt]
p_{va}\doteq SoftMax(s_{va}) \longrightarrow
\mathcal{L}_{va}\doteq CrossEntropy(p,p_{va})
\end{array}
\end{equation}

\item Network $\mathcal{N}_2(f_v,f_a;p)$ with normalized residual perceptron, fusion component and fused loss function 
$\mathcal{L} \doteq \mathcal{L}_v + \mathcal{L}_a +
\mathcal{L}_{va}$: 

\begin{equation}
\begin{array}{l}
\hat{f}_v\doteq FeatureNorm(f_v),\ 
\hat{f}_a\doteq FeatureNorm(f_a)\\[2pt]
f'_v\doteq ResPerceptron(\hat{f}_v),\ 
f'_a\doteq ResPerceptron(\hat{f}_a)\\[2pt]
s_v\doteq DenseUnit(\hat{f}_v),\ 
s_a\doteq DenseUnit(\hat{f}_a),\ 
s_{va} \doteq FusionComponent(f'_v,f'_a)
\\[2pt]
p_{v}\doteq SoftMax(s_{v}) \longrightarrow
\mathcal{L}_{v}\doteq CrossEntropy(p,p_{v})
\\[2pt]
p_{a}\doteq SoftMax(s_{a}) \longrightarrow
\mathcal{L}_{a}\doteq CrossEntropy(p,p_{a})
\\[2pt]
p_{va}\doteq SoftMax(s_{va}) \longrightarrow
\mathcal{L}_{va}\doteq CrossEntropy(p,p_{va})
\end{array}
\end{equation}
\end{enumerate}

\end{paracol}
\begin{figure}[ht]
\begin{center}
\includegraphics[width=.8\columnwidth]{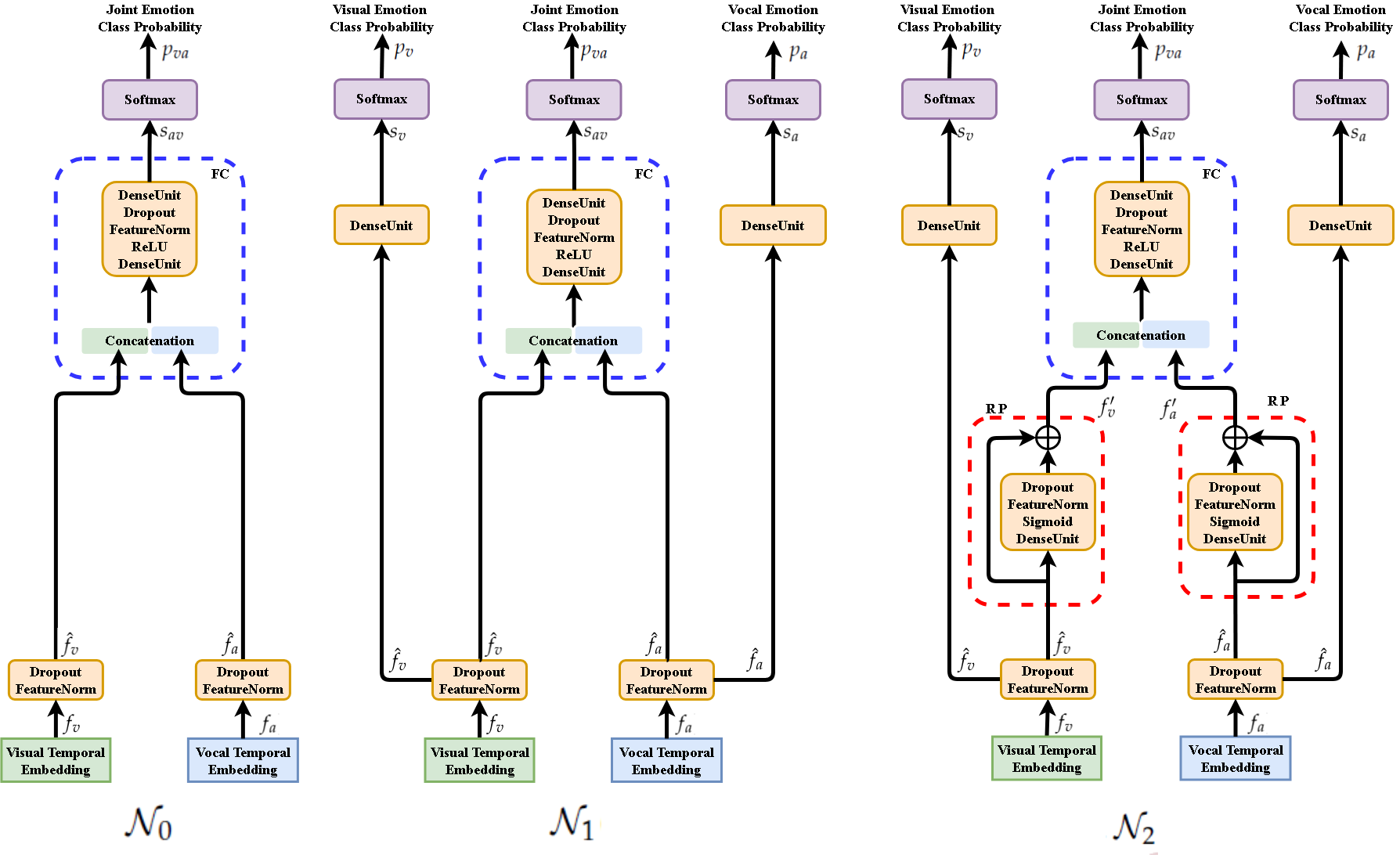}
\caption{Evolution of network design for multi-modal fusion (presented for training mode). $\mathcal{N}_0$: Fusion component (FC) only. $\mathcal{N}_1$ (\cite{9156420}): Beside FC, independent scoring of each modality is considered. $\mathcal{N}_2$: Extending $\mathcal{N}_1$ network by Residual Perceptrons (RP) in each modality branch.}
\label{details}
\end{center}
\end{figure}
\begin{paracol}{2}
\linenumbers
\switchcolumn

For the networks $\mathcal{N}_0, \mathcal{N}_1, \mathcal{N}_2$ detailed in Figure~\ref{details}, we can observe:

\begin{enumerate}
\item All instances of {\it FeatureNorm} unit are implemented as batch normalization units.
\item In testing mode only the central branch of networks $\mathcal{N}_1, \mathcal{N}_2$ are active while the side branches are inactive as they are used only to compute the extra terms of the extended loss function.

\item The above facts make network architectures $\mathcal{N}_0, \mathcal{N}_1$ equivalent in the testing mode. However, the models trained for those architectures are not the same, as weights are optimized for different loss functions. 

\item In the testing mode all {\it Dropout} units are not active, as well. 

\item The architecture of {\it FusionComponent} is identical for all three networks. The difference between models of $\mathcal{N}_0$ and $\mathcal{N}_1$ networks follows from the different loss functions while the difference between models of $\mathcal{N}_1$ and $\mathcal{N}_2$ networks is implied by using {\it ResPerceptron (RP)} components in $\mathcal{N}_2$ network.
\item To control the range of affine combinations computed by {\it Residual Perceptron (RP)} component, we use {\it Sigmoid} activations instead of the {\it ReLU} activations exploited in other components. The experiments confirm the advantage of this design decision.
\item The {\it Residual Perceptron (RP)} was introduced in the network $\mathcal{N}_2$ to implement better parameterization of within-modal features before their fusion.
\end{enumerate}
\subsection{MRPN components' role in multi-term optimization}
\begin{enumerate}

\item As we discussed in the hypothesis section, the late fusion strategy has the advantage of preserving the best information in each uni-modality, since the uni-modality extracts generalized deep features which suffer few from the outliers of their own modality. i.e. a small amount of wrongly labeled data in uni-modal solutions won't contribute to the generalized feature patterns, they are ``filtered out'' by the uni-modal neural network. Thus the additional term of loss functions implies the blended gradient in the shallow layers of each uni-modality, and helped for better parameterization of the features before fusion, preserving the knowledge as uni-modalities are trained respectively. The above facts make the end-to-end strategy suffer less inter-modal information as late fusion does. 

\item However, the multi-term optimization can result in extracting inferior uni-modal features as the input to the fusion component. This problem was mentioned in the literature \cite{standley2020tasks, 10.1023/A:1007379606734, Yu2020GradientSF}. RP is introduced to make modified uni-modal features, instead of storing all knowledge for uni-modal and multi-modal purposes in one unit, causing the clash of loss converging from two directions, the uni-modal and multi-modal knowledge can be stored in the original uni-modal features and modified multi-modal features, creating a new path for the gradient flow. RP can preserve the best of the uni-modal solution while the modified features from the short-cut can still fulfill the purpose of integrating new multi-modal features.
\end{enumerate}
The mentioned two novel properties make MRPN free from side-effects of late fusion and end-to-end strategy while preserving their own advantages. 
\subsection{MRPN in general multi-modal applications}

\begin{figure}[H]
\begin{center}
\includegraphics[width=0.6\columnwidth]{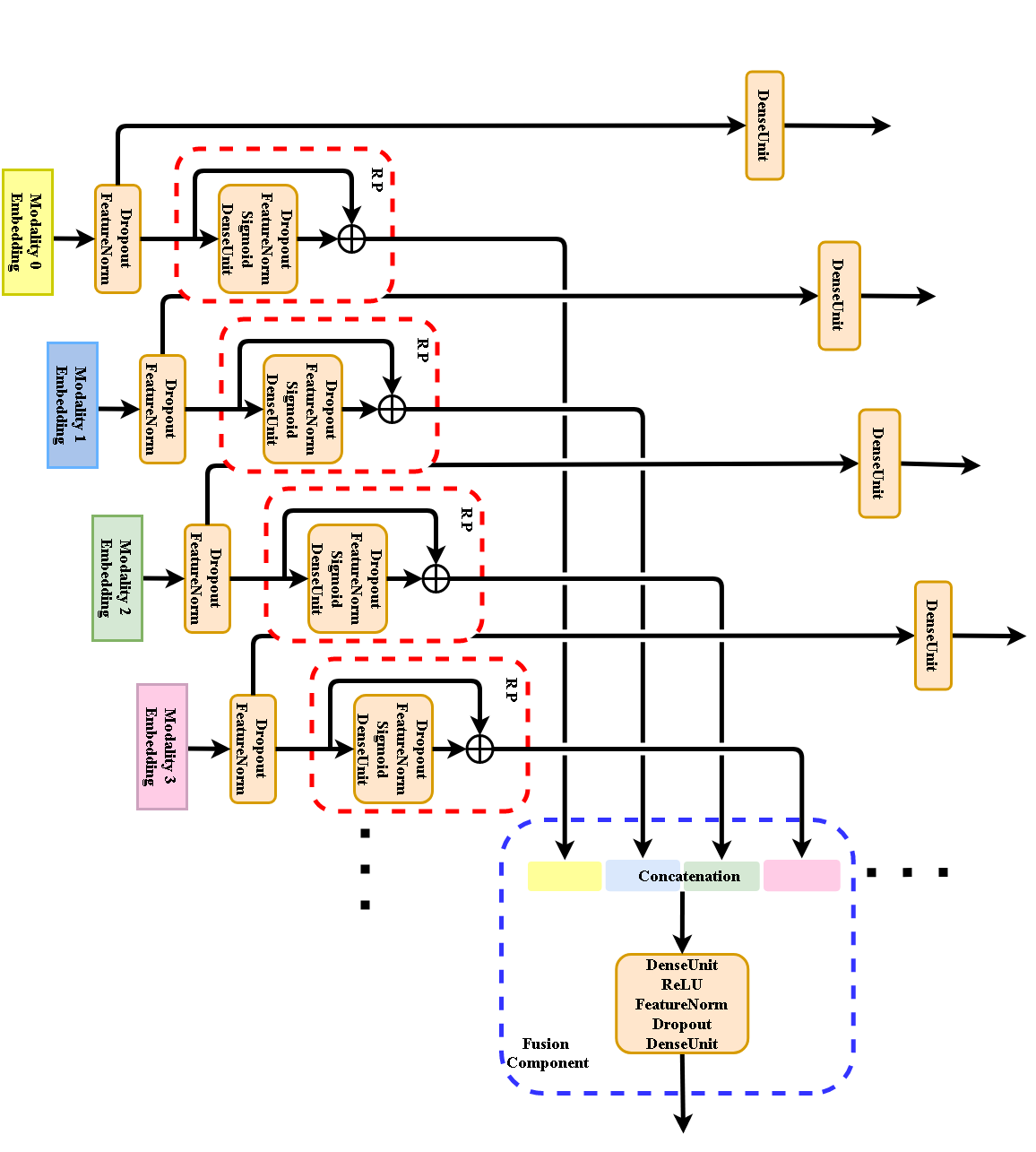}
\caption{Generalization of our MRPN fusion approach to many modalities. It could be used for either regression or classification applications.}
\label{generalMRPN}
\end{center}
\end{figure}
We suggest that MRPN can be adopted in any multi-modal application that involves many multi-modal inputs and one target function or many multi-modal inputs and many multi-modal target functions as Figure~\ref{generalMRPN} shows. In both cases MRPN benefits from many terms of loss functions as the numbers of uni-modalities, updating the whole system together while avoiding learning from inter-modal fuzzy information. MRPN is general to be compatible with any other proposed mechanism.
\subsection{Pre-processing}
Our data pre-processing includes the procedures for both modality inputs, namely spatial and time-dependent augmentation are applied.

\end{paracol}
\begin{figure}[ht]
\begin{center}
\includegraphics[width=.6\columnwidth]{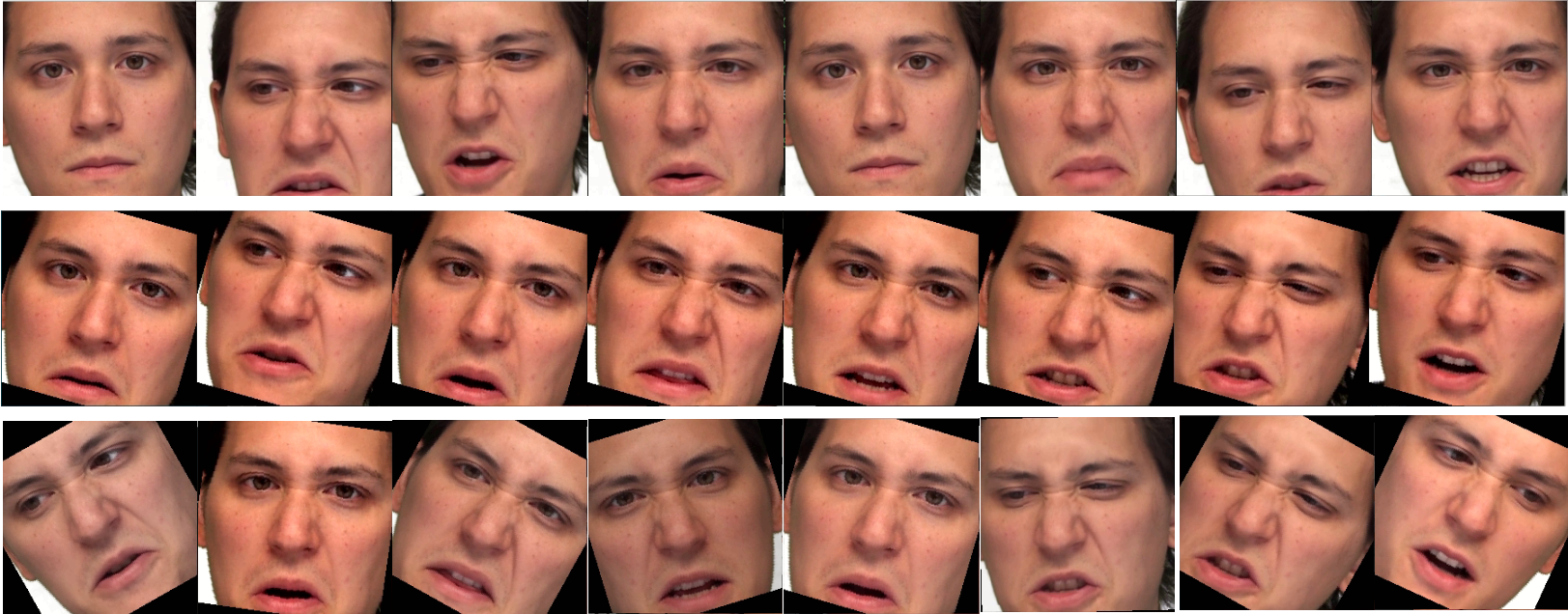}
\caption{\label{fig:emo_seq}Visual comparison of augmentation procedure for cropped video frames. \textbf{ Top part}: original video frames. \textbf{ Middle part}: applying random augmentation parameters -- same for all frames. \textbf{ Bottom part}: applying random augmentation parameters -- different for each frame.}
\end{center}
\end{figure}
\begin{paracol}{2}
\linenumbers
\switchcolumn
\begin{enumerate}
\item \textit{ Spatial data augmentation for visual frames:}

The facial area for the visual input frames is cropped using a CNN solution from Dlib library \cite{10.5555/1577069.1755843}. Once the facial area is cropped, spatial video augmentation is applied during the training phase. The same random augmentation parameters are applied for all frames of a video source illustrated in Figure~\ref{fig:emo_seq}. 
\item \textit{Time dependent data augmentation for visual frames:}\\
Obviously, expressions from the same category do not last the same duration. To make our system robust to the inconsistent duration of the emotion events, we perform data augmentation in time by randomly slicing the original frames as Figure~\ref{fig:spe_overlap} illustrates. Such operation should also avoid too few input frames missing information of the expression events. Thus the training segments are selected to have at least one-second duration unless the original duration of the file is less than that. 
\item \textit{Spatial data augmentation for vocal frames:}
Raw audio inputs are resampled at 16kHz and standardized by their mean and standard deviation without any denoising or cutting, to remove influence from the distance of the speaker to the microphone, or subjective base volume of the speaker. The standardized wave is then divided into one-second segments and converted to spectrograms by a Hann windowing function of size 512 and hop size of 64. 

The above facts specify the size of the spectrogram at 256 $\times$ 250, approaching the required input shape of Resnet-18 -- the CNN extractor used in our experiment. The chunk size using such inputs for Resnet-18 \cite{DBLP:journals/corr/HeZRS15} is close to its desired performance utilizing the advantage in the middle deep features. 

\item \textit{Time dependent augmentation for vocal frames:}
Similar to the time augmentation in visual inputs, raw audio inputs are also randomly sliced. The raw data is further over-sampled in both the training and testing mode by a hopping window. A window of 0.2 seconds, 1/5 duration of the input segments to the CNN extractor is specified. The oversampling further improved our results by the increasing number of deep features from the output deep feature sequences of CNN to the SAC. The mechanism grants the opportunity for the SAC to investigate more details of the temporal information in the deep feature vectors.
\end{enumerate}

\end{paracol}
\begin{figure}[ht]
\begin{center}
\includegraphics[width=.7\columnwidth]{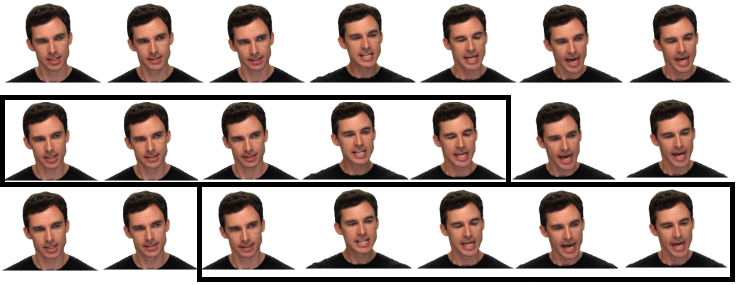}
\caption{\label{fig:spe_overlap}Examples of time dependent augmentation for visual frames. \textbf{ Top part}: original frames. {\it Middle part}: sliced frames start at beginning of the original frames. \textbf{ Bottom part}: sliced frames start at the middle of the original frames.}
\end{center}
\end{figure}
\begin{paracol}{2}
\linenumbers
\switchcolumn

\section{Computational Experiments and their Discussion}
This section presents an evaluation regarding the advantages of our proposed framework and time-dependent augmentation. Two datasets, RAVDESS and Crema-d are employed for this purpose. The improvement of the time-augmentation mechanism is analyzed in the naive fusion model which brought us state-of-art results even without MRPN design. The inferior cases in such common neural multi-modal solutions are detected and discussed in the comparison. Improvement of MRPN is then presented, not just in the detected inferior sub-datasets, but also in general data samples. 
\subsection{Datasets}
RAVDESS and Crema-d differ in numbers of expression categories, total files, identifies, and also video quality. 
\begin{enumerate}
\item RAVDESS dataset includes both speech and song files. For the speech recognition proposal, we only use the speech files from the dataset. It contains 2880 files, 24 actors (12 female, 12 male), state two lexically identical statements. Speech includes calm, happy, sad, angry, fearful, surprise, and disgusted expressions. Each expression is produced at two levels of emotional intensity (normal, strong), with an additional neutral expression, in a total of 8 categories. It is the most recent video-audio emotional dataset with the highest video quality in this research area to our best knowledge.

\item Crema-d dataset consists of visual and vocal emotional speech files in a range of basic emotional states (happy, sad, anger, fear, disgust, and neutral). 7,442 clips of 91 actors with diverse ethnic backgrounds were rated by multiple raters in three modalities: audio, visual, and audio-visual.
\end{enumerate}

For both datasets, the training set and testing set are separated using similar concepts as 10-fold inter-validation. Additionally, identities of the actors are also separated in train and val sets to prevent the results from leaning on the actors. Around 10\% of the actors are used for validation while the remaining 90\% are used for training, male and female actors are balanced in each set. We rotate the split train/validation sub-datasets to get multiple results over the whole dataset. 
The crema-d dataset has fewer categories for the classification tasks, but from the report of the authors themselves, Crema-d holds the accuracy of 63.6\% from human recognition for 6 categories which are less than RAVDESS at 72.3\% for 8 categories. The resolution of the video source is verified not the cause of the worse performance. The better results in the RAVDESS dataset, in our opinion, are the more crystal and natural emotion information inside the RAVDESS dataset.

\subsection{Model organization and computational setup}
The naive fusion model $\mathcal{N}_0$, advanced fusion network $\mathcal{N}_1$, which is equivalent to the Facebook \cite{9156420} solution and the $\mathcal{N}_2$ (MRPN) have the same CNN extractors at the initial stage of the training. To compare the impact of strategy from features fusion only, CNN extractor architecture is fixed to Resnet-18 \cite{DBLP:journals/corr/HeZRS15}. 

The CNN in visual modality is initialized from a facial image expression recognition task, the challenge FER2013 \cite{GOODFELLOW201559}. As for vocal modality, The CNN is pretrained on the voice recognition task from VoxCeleb dataset \cite{Nagrani_2017}. The initialization of the CNN extractors made the whole system much easier to be optimized. 

AdaMW optimizer is adopted for the model optimization, with the initial learning rate at $5\cdot 10^{-5}$, decreased two times if validation loss is not dropping over ten epochs.
\subsection{Data augmentation cannot generalize multi-modal feature patterns}

This subsection illustrates the improvement of time-dependent augmentation. The improvement also proves that the inferior case of multi-modal solution doesn't depend on the with-modal patterns. The single modality solutions in our experiments (shown in Table~\ref{tacompare}) take pretrained Resnet-18 as extractors and LSTM cells as SACs. The naive multi-modal solution takes twice of the components with an additional fusion layer as Figure~\ref{details} illustrates on the left panel. Adopting time-dependent augmentation shows overall performance improvements on either single or multi-modal solutions.
The Table notations are presented in the follows:

In the variational train/val sub-datasets in Table~\ref{tacompare}, Ax,y stands for the validation files that came from actor x and y, odd number notes for a male actor, and even number for a female actor.

\begin{specialtable}[ht] 
\caption{Comparison of single modalities models with $\mathcal{N}_0$ model (RAVDESS cases): VM -- Visual Modality only, AM -- Audio Modality only, JM -- Joint Modalities ($\mathcal{N}_0$ model), T -- having time augmentation by signal random slicing, NT -- not having time augmentation.\label{tacompare}}
\begin{tabular}{ccccccc}
\toprule
RAVDESS & A1,2   & A3,4   & A5,6   & A7,8   & A9,10  & A11,12 \\ 
\midrule
AM (NT)  & 70.8\% & 55.0\% & 57.5\% & 74.1\% & 43.5\% & 65.8\% \\
AM (T)   & 71.6\% & 77.5\% & 71.6\% & 90.0\% & 55.8\% & 69.1\% \\ 
\midrule
VM (NT)  & 82.5\% & 70.0\% & 66.7\% & 74.1\% & {\bf80.3}\% & 63.3\% \\
VM (T)   & 86.6\% & 75.0\% & 70.6\% & 76.6\% & {\bf 87.3\%} & 69.1\% \\ 
\midrule
JM (NT)  & 90.8\% & 89.1\% & 85.2\% & 89.3\% & {\bf78.5}\% & 85.5\% \\
JM (T)   & 97.5\% & 90.3\% & 87.5\% & 97.5\% & {\bf86.5}\% & 87.5\% \\ 
\midrule
RAVDESS & A13,14 & A15,16 & A17,18 & A19,20 & A21,22 & A23,24 \\ 
\midrule
AM (NT)  & 59.8\% & 57.5\% & 51.6\% & 55.5\% & 55.8\% & 63.3\% \\ 
AM (T)   & 70.0\% & 69.1\% & 57.5\% & 63.3\% & 68.3\% & 68.3\% \\ 
\midrule
VM (NT)  & 71.3\% & 60.0\% & 63.3\% & 70.8\% & 65.8\% & 70.8\% \\ 
VM (T)   & 73.3\% & 65.0\% & 64.1\% & 78.3\% & 66.6\% & 74.1\% \\ 
\midrule
JM (NT)  & 77.5\% & 75.5\% & 76.3\% & 85.2\% & 82.8\% & 80.0\% \\ 
JM (T)   & 82.4\% & 79.6\% & 83.2\% & 89.0\% & 85.5\% & 84.2\% \\ 
\bottomrule
\end{tabular}
\end{specialtable}

\subsection{Discussion on inferior multi-modal cases }
The time augmentation shows overall improvements in either uni-modal or multi-modal approach, yet the inferior case where uni-modal solution better than multi-modal solution still exists, which suggests data augmentation cannot generalize multi-modal features.
Only one inferior case is detected in Table~\ref{tacompare} of the case A9,10,
but we argue such deficiency is common in fuzzy multi-modal data. The pattern learning ability from both modalities is well enough, both solutions have performance over 85\% in cases like A7,8 and A1,2. But the ratio of mismatched learned and target patterns are ranging along with the shuffling of the sub-datasets. 

The degeneration of the performance became visible only because the percentage of the pattern mismatched samples has passed some kind of threshold in the training set. If so, by eliminating or reducing such side effects, overall improvements should be expected for any train and testing sub-dataset. 
\subsection{Improvement of MRPN}
This subsection addresses the improvement of MRPN preventing the side-effects in the existing late fusion and end-to-end strategies we hypothesized as Table~\ref{MRPN1} and Table~\ref{MRPN2} illustrate.
%
The end-to-end strategy of $\mathcal{N}_1$, which takes multi-term loss function helped the better parameterization shows improved average performance over naive end-to-end and late fusion training strategies, yet it can still fail in some cases. Our proposed MRPN on the contrary demonstrates the same performance or most improvement in any circumstance.

\begin{specialtable}[H] 
\caption{Comparison for RAVDESS of MRPN approach (network $\mathcal{N}_2$) with late fusion strategy ($\mathcal{N}_0$), end-to-end strategy ($\mathcal{N}_0$), and advanced end-to-end fusion strategy ($\mathcal{N}_1$).\label{MRPN1}}
\begin{tabular}{lcccccc}
\toprule
RAVDESS	& A1,2   & A3,4   & A5,6   & A7,8   & A9,10  & A11,12\\
\midrule
$\mathcal{N}_0$(late fusion)&61.6\%&92.1\%&87.5\%&96.6\%&66.6\%&87.5\% \\
$\mathcal{N}_0$(end-to-end) & \textbf{97.5}\% & 90.3\% & 87.5\% & \textbf{97.5}\% & 86.5\% & 87.5\% \\
$\mathcal{N}_1$(end-to-end)      & \textbf{97.5}\% & 89.1\% & 88.3\% & \textbf{97.5}\% & 90.0\% & \textbf{90.0}\% \\ 
$\mathcal{N}_2$(end-to-end)         & \textbf{97.5}\% & \textbf{92.1}\% & \textbf{90.8}\% & \textbf{97.5}\% & \textbf{91.4}\% & \textbf{90.0}\% \\ 
\midrule
RAVDESS       & A13,14 & A15,16 & A17,18 & A19,20 & A21,22 & A23,24 \\ 
\midrule
$\mathcal{N}_0$(late fusion)&80.8\%&85.0\%&81.6\%&87.5\%&86.6\%&65.8\% \\
$\mathcal{N}_0$(end-to-end)       & 82.4\% & 79.6\% & 83.2\% & 89.0\% & 85.5\%    & 84.2\%    \\ 
$\mathcal{N}_1$(end-to-end)        & 77.5\% & 89.1\% & 86.6\% & 92.5\% & 89.1\% & \textbf{90.6}\% \\ 
$\mathcal{N}_2$(end-to-end)           & \textbf{84.3}\% & \textbf{89.7}\% & \textbf{89.8}\% & \textbf{93.3}\% & \textbf{90.6}\%    & \textbf{90.6}\%    \\
\bottomrule
\end{tabular}
\end{specialtable}

\end{paracol}
\begin{figure}[ht]
\begin{center}
\includegraphics[width=.24\columnwidth]{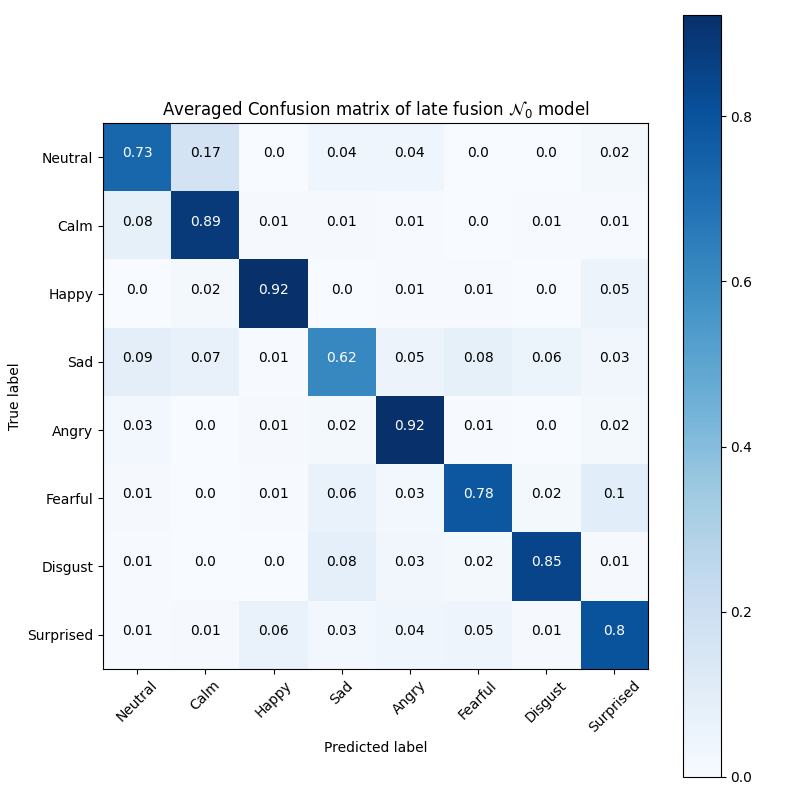}
\includegraphics[width=.24\columnwidth]{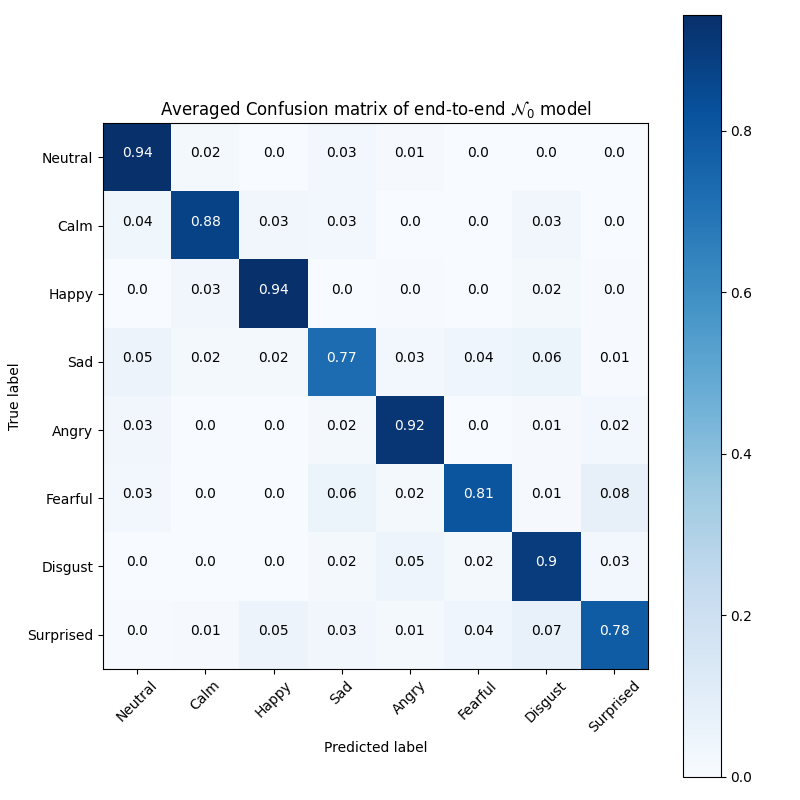}
\includegraphics[width=.24\columnwidth]{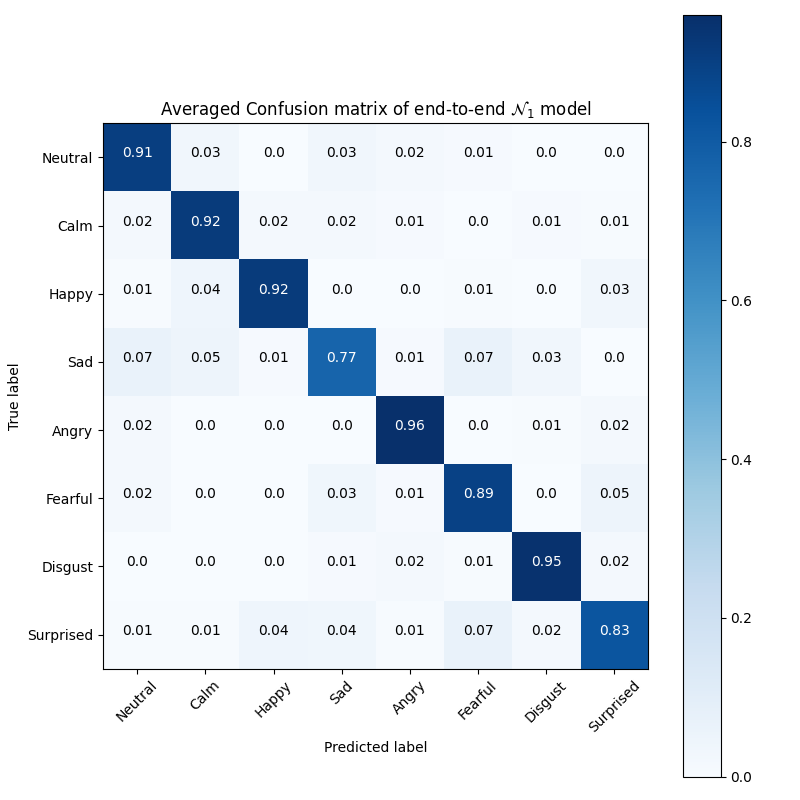}
\includegraphics[width=.24\columnwidth]{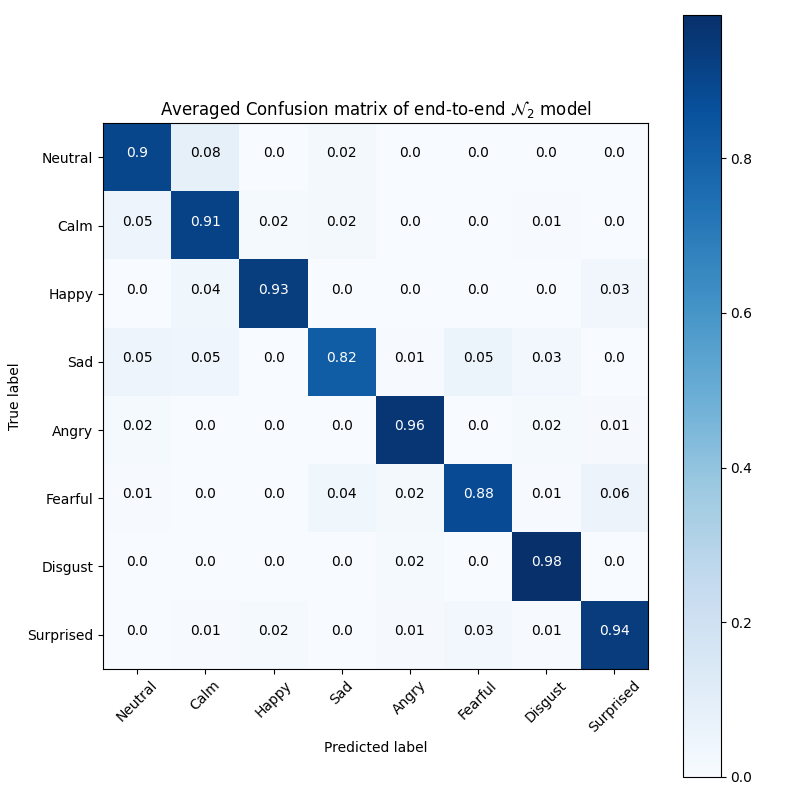}
\caption{Averaged confusion matrices of tested models for RAVDESS dataset.\label{cm1}}
\end{center}
\end{figure}
\begin{paracol}{2}
\linenumbers
\switchcolumn

It can been seen from the confusion matrices in Figure~\ref{cm1} and \ref{cm2} the averaged improvements of $\mathcal{N}_2$ (MRPN) over the late fusion and end-to-end $\mathcal{N}_0$ models. Performance on some specific categories shows a slight decrease for MRPN, especially for the categories of calm and neutral expressions because they are naturally close to each other in the RAVDESS dataset. $\mathcal{N}_1$ doesn't always perform better than the existing solutions, the almost 6\% improvements of $\mathcal{N}_2$ (MRPN) over $\mathcal{N}_1$ suggests the level of data fuzziness can make the end-to-end multi-term optimization even harder without proposed RP components. The overall improvements suggest that multi-modal patterns are more generalized from the solution of $\mathcal{N}_2$ (MRPN).

\begin{specialtable}[H] 
\caption{Comparison for Crema-d of MRPN approach (network $\mathcal{N}_2$) with simple fusion strategy ($\mathcal{N}_0$), and advanced fusion strategy ($\mathcal{N}_1$).\label{MRPN2}}
\begin{tabular}{lccccc}
\toprule
Crema-d       & S1     & S2     & S3     & S4     & S5     \\ 
\midrule
$\mathcal{N}_0$ (late fusion)&76.5\%&79.9\%&76.6\%&62.3\%&78.2\%\\
$\mathcal{N}_0$ (end-to-end) & 77.3\% & 81.3\% & 79.2\% & 74.8\% & 78.6\% \\
$\mathcal{N}_1$ (end-to-end)     & 72.6\% & 82.3\% & 77.3\% & 74.8\% & 74.2\%  \\ 
$\mathcal{N}_2$ (end-to-end)               & \textbf{79.5}\% & \textbf{83.0}\% &\textbf{ 83.0}\% & \textbf{76.8}\% & \textbf{81.9}\% \\ 
\midrule
Crema-d       & S6     & S7     & S8     & S9     &                             \\ 
\midrule
$\mathcal{N}_0$ (late fusion) &81.8\%&78.8\%&80.0\%&77.5\%\\
$\mathcal{N}_0$ (end-to-end) &\textbf{ 82.0}\% & 75.1\% &79.5\% &77.5\%                            \\
$\mathcal{N}_1$ (end-to-end)      &\textbf{82.0}\% & 74.8\% &79.3\% &75.8\% \\ 
$\mathcal{N}_2$ (end-to-end)               &\textbf{82.0}\% &\textbf{80.0}\% &\textbf{80.5}\% &\textbf{78.6}\%                             \\ 
\bottomrule
\end{tabular}
\end{specialtable}

\end{paracol}
\begin{figure}[ht]
\begin{center}
\includegraphics[width=.24\columnwidth]{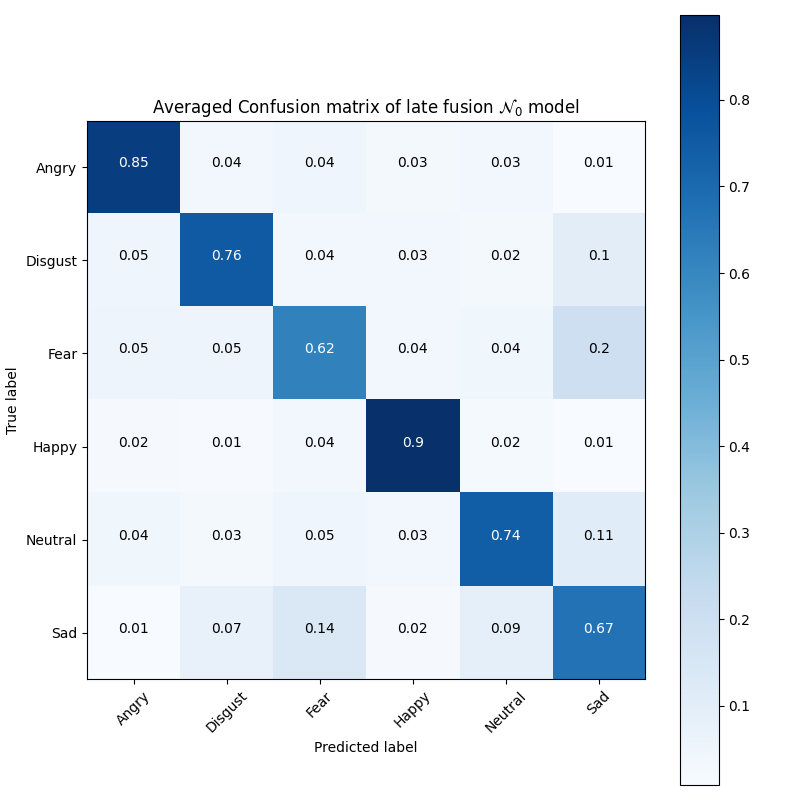}
\includegraphics[width=.24\columnwidth]{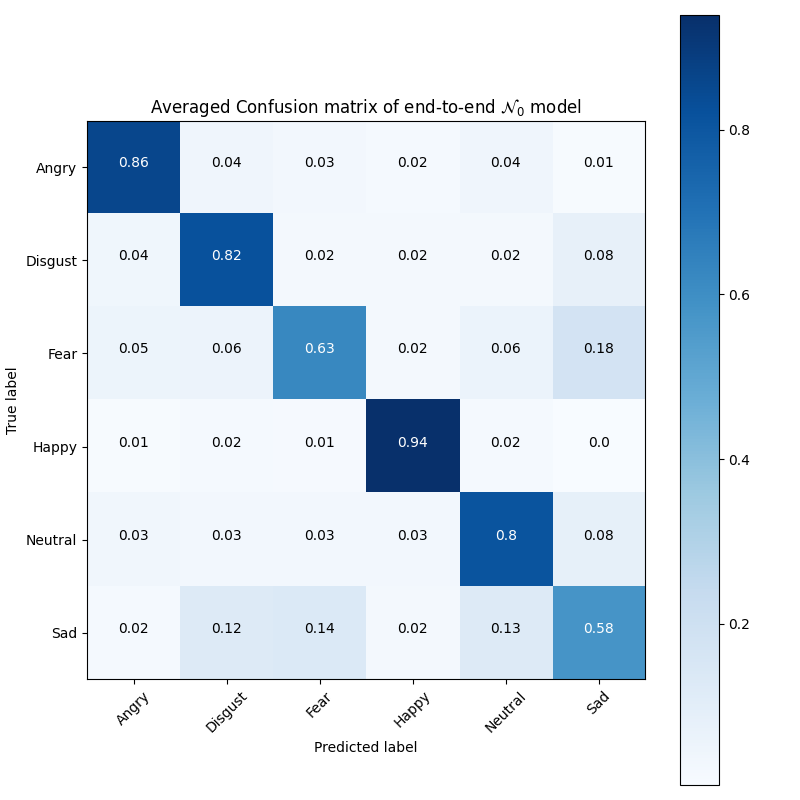}
\includegraphics[width=.24\columnwidth]{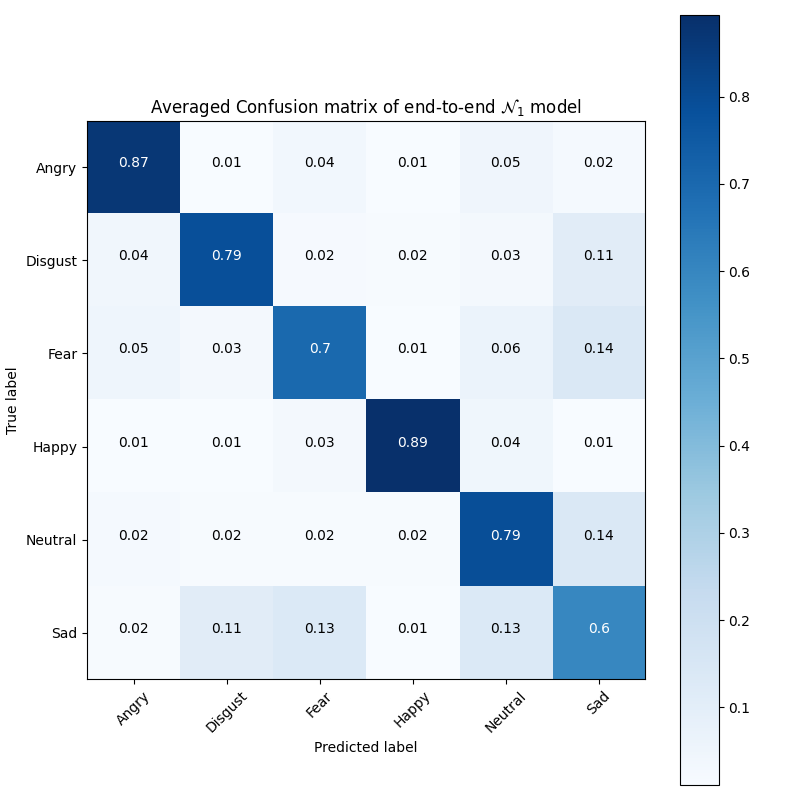}
\includegraphics[width=.24\columnwidth]{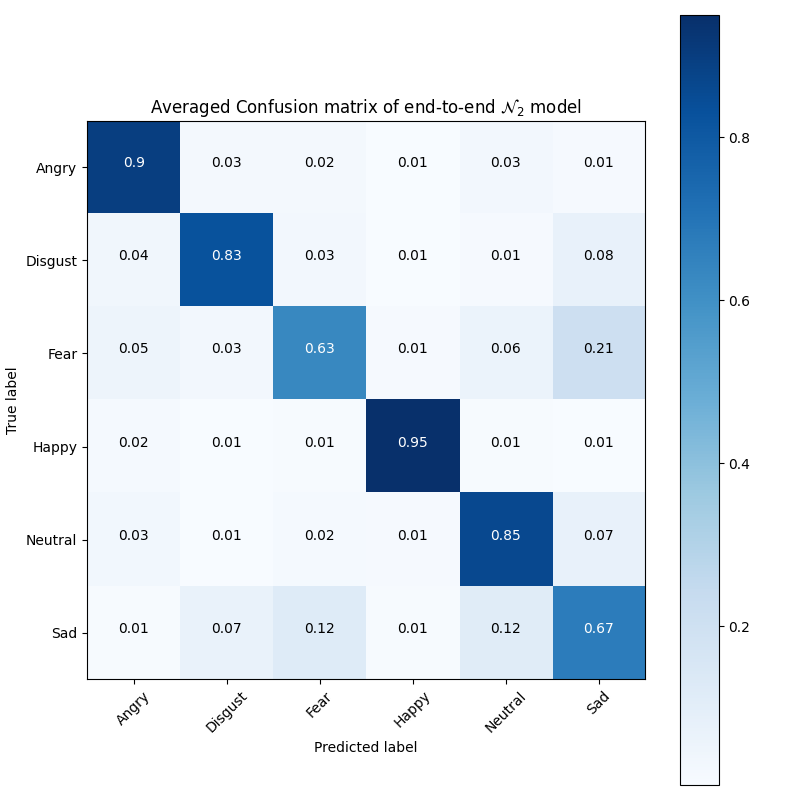}
\caption{Averaged confusion matrices of tested models for Crema-d dataset.\label{cm2}}
\end{center}
\end{figure}
\begin{paracol}{2}
\linenumbers
\switchcolumn
\subsection{Comparing baseline with SOTA}
Our proposed MRPN shows stat-or-art results on both datasets. It has no conflicts with any potential advantages from another novel mechanism. Experiments regarding the pretraining of the CNN extractors and the time augmentation have made the network robust to overcome the overfitting issues regarding the small amount of the training and testing data. 

Additionally, replacement of LSTM of Bidirectional LSTM and Transformer as aggregator has been conducted but no noticeable differences of them as sequence aggregators can be seen. As for Transformer, the average feature is taken from the decoded outputs, the concept follows from Vision Transformer (VIT) \cite{dosovitskiy2020image}.

\begin{specialtable}[H] 
\caption{Comparison of our fusion models with others recent solutions. Options used: IA -- image augmentation, WO -- without audio overlapping, VA -- video frames augmentation, and AO -- audio overlapping. X symbol -- there is no report from authors for the given dataset.\label{compare}}
\begin{tabular}{lcc}
\toprule
\textbf{Model (our)}	& \textbf{RAVDESS}	& \textbf{Crema-d}\\
\midrule

$\mathcal{N}_0$ (end-to-end),Resnet18+LSTM, IA                                         & 83.20\%  & 77.25\%  \\
$\mathcal{N}_0$ (end-to-end),Resnet18+LSTM, VA+WO                                            & 85.20\%  & 79.25\%  \\
$\mathcal{N}_0$ (late fusion),Resnet18+LSTM, VA+AO                                             & 81.6\%  & 76.84\%  \\
$\mathcal{N}_0$ (end-to-end),Resnet18+LSTM, VA+AO                                             & 87.55\%  & 81.30\%  \\
$\mathcal{N}_1$  (end-to-end),Resnet18+LSTM, VA+AO  \cite{9156420}                                           & 89.8\%  &77.0\%  \\
MRPN (end-to-end),Resnet18+LSTM, VA+AO                                                   & \bf{90.8\%}  & \bf{83.00\%}  \\
MRPN (end-to-end),Resnet18+Transformer(avg), VA+AO                                                   & \bf{91.4\%}  & \bf{83.15\%}  \\
\toprule
\textbf{Model (others)}	& \textbf{RAVDESS}	& \textbf{Crema-d}\\
\midrule
(OpenFace/COVAREP features + LSTM) + Attention \cite{beard-etal-2018-multi} & 58.33\% & 65.00\%      \\
Dual Attention + LSTM \cite{8925444}               & 67.7\%  & 74.00\%  \\
Resnet101 + BiLSTM \cite{9078789}                       & 77.02\% &   X      \\ 
custom CNN \cite{8906538}                               &     X    & 69.42\% \\
Early Cross-modal + MFCC + MEL spectrogram \cite{8852473} & 83.6\%&X \\
CNN + Fisher vector + Metric learning \cite{8935376}&X&66.5\%\\
custom CNN+Spectrogram \cite{s20010183}& 79.5\% (Audio)&X\\
\bottomrule
\end{tabular}
\end{specialtable}

\section{Conclusion}
This paper focuses on explaining the potential deficiencies in the existing fusion layer of the multi-modal approach to AVER tasks using late fusion or end-to-end strategy.
The proposed MPRN architecture along with the multi-term loss function makes superior fused features from multi-modal sources. We observe the elimination of inferior cases of multi-modal solutions with respect to uni-modal solutions.

Our results achieve an average accuracy of 91.4\% on the RAVDESS dataset and 83.15\% on the Crema-d dataset. MRPN solution contributes to a better
 average recognition rate of approximately 2\%. We have observed the maximum improvement of MRPN for a subset to be around 90\% from nearly 80\%. 

The proposed data pre-processing by time augmentation makes general overall rate improvements for both, the uni-modal and multi-modal data. It also illustrates data augmentation cannot generalize multi-modal features due to the deficiencies in the existing multi-modal solutions.

Moreover, the MRPN concept shows its potential for multi-modal classifiers dealing with signal sources not only of optical and acoustical type.





\authorcontributions{Conceptualization, X.C. and W.S.; Data curation, X.C.; Formal analysis, X.C. and W.S.; Investigation, X.C. and W.S.; Methodology, X.C. and W.S.; Project administration, X.C. and W.S.; Resources, X.C. and W.S.; Software, X.C. and W.S.; Supervision, W.S.; Validation, X.C. and W.S.; Visualization, X.C. and W.S.; Writing – original draft, X.C. and W.S.; Writing – review \& editing, X.C. and W.S.}
\funding{Beside statutory support of Warsaw University of Technology this research received no external funding.}

\institutionalreview{Not applicable.}

\informedconsent{Not applicable.}

\dataavailability{The initialization of visual CNN extractor were sampled from the FER-2013 challenge \cite{GOODFELLOW201559}. This data
was made available under the Open Database License and can be found at \url{https://www.kaggle.com/msambare/fer2013}. (accessed on 7 July 2021). The initialization of vocal CNN extractor were sampled from VoxCeleb dataset\cite{Nagrani_2017}. This data
was made available under the Open Database License and can be found at \url{https://www.robots.ox.ac.uk/~vgg/data/voxceleb/vox1.html}. (accessed on 7 July 2021).
The evaluation of the MRPN and time augmentation is based on RAVDESS dataset\cite{10.1371/journal.pone.0196391} and Crema-d\cite{6849440} under the Open Database License and can be found at \url{https://zenodo.org/record/1188976\#.YOOv3Oj7SUk}(accessed on 7 July 2021) and \url{https://github.com/CheyneyComputerScience/CREMA-D}(accessed on 7 July 2021) respectively.} 


\conflictsofinterest{The authors declare no conflict of interest.} 

\abbreviations{Abbreviations}{
\noindent 
\begin{tabular}{@{}ll}
AVER&Audio-video emotion recognition \\
CNN&Convolution Neural Network\\
Crema-d \cite{6849440}&Crowd-sourced Emotional multi-modal Actors Dataset\\
DNN&Deep Neural Network\\
HCI&Human-Computer Interaction\\
FC&FC Fully connected layers\\
IEMOCAP \cite{iemocap}&Interactive emotional dyadic motion capture dataset\\ 
LSTM&Long Short Term Memory\\
MRPN&multi-modal Residual Perceptron Network\\
RAVDESS \cite{10.1371/journal.pone.0196391}&The Ryerson Audio-Visual Database of Emotional Speech and Song \\
RP&Residual Perceptron\\
SAC&Sequence Aggregation Component\\
SOTA & State of the Art Solution\\
STFT& Short-term Fourier transformation\\
SVM&Support Vector Machine\\
VIT \cite{dosovitskiy2020image}&Vision Transformer\\
\end{tabular}}



\end{paracol}
\reftitle{References}


\externalbibliography{yes}

\bibliography{referred} 
%


\end{document}